# Monitoring the carbon emissions transition of global building end-use activity[#]

Xiwang Xiang [1], Minda Ma [2*]

1 School of Management Science and Real Estate, Chongqing University, Chongqing, 400045, PR China

2 Building Technology & Urban Systems Division, Lawrence Berkeley National Laboratory, Berkeley, CA 94720, United States
(Corresponding Author, maminda@lbl.gov; minda.ma@cqu.edu.cn)

**ABSTRACT**

The building sector is the largest emitter globally and as such is at the forefront of the net-zero emissions pathway. This study is the first to present a bottom-up assessment framework integrated with the decomposing structural decomposition method to evaluate the emission patterns and decarbonization process of global residential building operations and commercial building operation simultaneously over the last two decades. The results reveal that (1) the average carbon intensity of global commercial building operations has maintained an annual decline of 1.94% since 2000, and emission factors and industrial structures were generally the key to decarbonizing commercial building operations; (2) the operational carbon intensity of global residential buildings has maintained an annual decline of 1.2% over the past two decades, and energy intensity and average household size have been key to this decarbonization; and (3) the total decarbonization of commercial building operations and residential buildings worldwide was 230.28 and 338.1 mega-tons of carbon dioxide per yr, respectively, with a decarbonization efficiency of 10.05% and 9.4%. Overall, this study assesses the global historical progress in decarbonizing global building operations and closes the relevant gap, and it helps plan the stepwise carbon neutral pathway of future global buildings by the mid-century.

**Keywords:** global building sector, operational carbon emissions, decomposing structural decomposition, building decarbonization.

**NONMENCLATURE**

| | |
|---|---|
| *Abbreviations* | |
| DSD | Decomposing Structural Decomposition |
| GDP | Gross Domestic Product |
| GtCO$_2$ | Gigatons of carbon dioxide |
| kgCO$_2$ | Kilograms of carbon dioxide |
| MtCO$_2$ | Million tons of carbon dioxide |
| *Symbols* | |
| $C^{res}, C^{com}$ | Operational carbon emission in residential and commercial buildings |
| $E^{res}, E^{com}$ | Energy consumption in residential and commercial buildings |
| F | Floor space of commercial buildings |
| GDP | Gross Domestic Product |
| Gs | GDP of service industry |
| H | Amounts of households |
| HFC | Household final consumption |
| P | Population size |

## 1. INTRODUCTION

The climate goal set out in the Paris Agreement to keep global warming well below 2 degrees and work toward 1.5 degrees leaves limited emission space for human activity and will be impossible to achieve without immediate and high decarbonization in all emission sectors [1]. The building sector is by far the world's largest energy consumer and emitter, with operational carbon from buildings reaching another record high of 10 gigatons CO$_2$ (GtCO$_2$) in 2021, 5% higher than when the Paris Agreement was signed in 2015 [2]. Therefore, decarbonization of global building sector is a crucial step toward achieving building carbon neutrality and the Paris climate goals.

To fast-track carbon neutrality in global building operations, nearly 160 countries worldwide have set ambitious reduction targets and have included building decarbonization as part of their nationally determined contributions [3]. Although these mitigation efforts have raised the national ambition level and provided guidance for a low-carbon transition in residences, not every country has contributed equally, and these efforts have focused primarily on the supply side (e.g., improving material efficiency [4] and energy-fuel decarbonization [5]). Indeed, there is growing evidence that end-use decarbonization will be fundamental to achieving this goal in the foreseeable future [6]. However, few studies have reviewed and assessed the decarbonization of residential building operations by region or country worldwide to determine the baseline of historical decarbonization and the potential for high decarbonization, especially a detailed assessment of the mitigation potential for measures related to end uses (e.g., space heating, appliances) [7]. Therefore, to address these gaps, this study proposes the following three questions for the global building sector (including commercial building and residential building):

- How has the operational carbon of commercial buildings changed since 2000?
- How has the operational carbon of residential buildings changed since 2000?
- What is the operational decarbonization level and how can its process be accelerated?

To answer these three questions, a bottom-up framework is developed to assess the historical decarbonization of global building operations over the past two decades. Decomposing structural decomposition (DSD) is integrated to identify the effects of the socio-economy, technological innovation, and end-use activity on operational decarbonization of building sector. Furthermore, the historical process and decarbonization efficiency of global buildings are assessed, and the historical decarbonization performance across regions is investigated and compared based on three emission scales.

**As its most important contributions,** this study aims to be the first to assess and compare global and regional historical decarbonization and the corresponding potential of building operations and to provide a benchmark for countries to fairly set decarbonization targets and remaining emission quotas. The high decarbonization of these major emitters will also inform the development of the remaining emerging emitting economies and reserve more emission space. To this end, this study estimates the spatial-temporal evolution patterns of the carbon intensity (i.e., carbon emissions per household and carbon emissions per floor space) in global residential building operations and commercial building operations and investigates the effects of socio-economy, technological innovation, and end uses on operational decarbonization worldwide, particularly measuring the decarbonization benefits across ten different end-use activities by considering the end-use structure and the corresponding emission factor change.

## 2. MATERIAL AND METHODS

### 2.1 Global Building emission model

Considering the comparability of end-use energy across different countries and better analyzing the impact of end-use activities on the decarbonization of global building operations, the end-uses of commercial buildings were decomposed into space heating, space cooling, service lighting, and appliances and others, and the end-uses of commercial buildings were decomposed into space heating, space cooling, service lighting, and appliances and others [8]. Thus, the global building emissions model can be defined as:

$$\begin{cases} C^{com} = C^{com}_{\text{Space heating}} + C^{com}_{\text{Space cooling}} + C^{com}_{\text{Service lighting}} + C^{com}_{\text{Appliances and other}} \\ C^{res} = C^{res}_{\text{Space heating}} + C^{res}_{\text{Space cooling}} + C^{res}_{\text{Water heating}} + C^{res}_{\text{Lighting}} + C^{res}_{\text{Cooking}} + C^{res}_{\text{Appliances \& others}} \end{cases} \quad (1)$$

To evaluate the operational carbon emissions, some potential influencing factors were taken into account to characterize the building emission model, including average household size ($\frac{P}{H}$), population density ($\frac{P}{F}$), industrial structure ($\frac{Gs}{GDP}$), economic efficiency ($\frac{F}{Gs}$), gross domestic product (GDP) per capita ($\frac{GDP}{P}$), household consumption capacity ($\frac{HFC}{GDP}$), energy intensity ($\frac{E}{HFC}$), end-use structure ($\frac{E_j}{E}$), and emission factors ($\frac{X_j}{E_j}$). Hence, Eq. (1) can be characterized as follows:

$$\begin{cases} c_j^{com} = \frac{C_j^{com}}{F} \equiv \frac{P}{F} \cdot \frac{GDP}{P} \cdot \frac{G_s}{GDP} \cdot \frac{F}{G_s} \cdot \frac{E_j^{com}}{F} \cdot \frac{C_j^{com}}{E_j^{com}} \\ c_j^{res} = \frac{C_j^{res}}{H} \equiv \frac{P}{H} \cdot \frac{GDP}{P} \cdot \frac{HFC}{GDP} \cdot \frac{E}{HFC} \cdot \frac{E_j^{res}}{E} \cdot \frac{C_j^{res}}{E_j^{res}} \end{cases} \quad (2)$$

For convenience, let $pc = \frac{P}{F}, pr = \frac{P}{H}, gc = \frac{GDP}{P}, gr = \frac{GDP}{P}, sc = \frac{G_s}{GDP}, hr = \frac{HFC}{GDP}, ic = \frac{F}{G_s}, er = \frac{E}{HFC}, s_j^{res} = \frac{E_j^{res}}{E}, k_j^{res} = \frac{C_j^{res}}{E_j^{res}}, e_j^{com} = \frac{E_j^{com}}{F}, k_j^{com} = \frac{C_j^{com}}{E_j^{com}}$, then, combining Eq. (1) and Eq. (2), the operational carbon intensity is defined as:

$$\begin{cases} c^{com} = \sum_{j=1}^{4} pc \cdot gc \cdot sc \cdot ic \cdot e_j^{com} \cdot k_j^{com} \\ c^{res} = \sum_{j=1}^{6} pr \cdot gr \cdot hr \cdot er \cdot s_j^{res} \cdot k_j^{res} \end{cases} \quad (3)$$

### 2.2 Decomposing structural decomposition method

Here, the DSD method was utilized to decompose the established emission model. Following the computational procedure of DSD, Eq. (3) can be expanded into the full differential equation as:

$$\begin{cases} dc^{res} = \frac{\partial c}{\partial pr} dpr + \frac{\partial c}{\partial gr} dgr + \frac{\partial c}{\partial hr} dhr + \frac{\partial c}{\partial er} der + \sum_{i=1}^{6} \left( \frac{\partial c_j}{\partial s_j^{res}} ds_j^{res} + \frac{\partial c_j}{\partial k_j^{res}} dk_j^{res} \right) \\ dc^{com} = \frac{\partial c}{\partial pc} dpc + \frac{\partial c}{\partial gc} dgc + \frac{\partial c}{\partial sc} dsc + \frac{\partial c}{\partial ic} dic + \sum_{i=1}^{4} \left( \frac{\partial c_j}{\partial e_j^{com}} de_j^{com} + \frac{\partial c_j}{\partial k_j^{com}} dk_j^{com} \right) \end{cases} \quad (4)$$

Guided by previous studies [9], the change in carbon intensity of global building operations can be decomposed by DSD as:

$$\begin{cases} \Delta c^{com}|_{0 \to T} = \Delta pc + \Delta gc + \Delta sc + \Delta ic + \Delta e^{com} + \Delta k^{com} \\ \Delta c^{res}|_{0 \to T} = \Delta pr + \Delta gr + \Delta hr + \Delta er + \Delta s^{res} + \Delta k^{res} \end{cases} \quad (5)$$

### 2.3 Operational decarbonization assessment

Based on the decomposition outcomes mentioned above, the decarbonization intensity for global buildings (i.e., decarbonization per household $DCI^{res}$ or decarbonization per floor space $DCI^{com}$) can be further assessed by the detrimental effects of factors driving the operational carbon intensity [10], as follows:

$$\begin{cases} DCI^{com}|_{0\to T} = -\sum(\Delta C_i^{com}|_{0\to T}) \\ DCI^{res}|_{0\to T} = -\sum(\Delta C_i^{res}|_{0\to T}) \end{cases} \quad (6)$$

where $\Delta C_i^{res} \in \{\Delta p, \Delta g, \Delta h, \Delta e, \Delta s^{res}, \Delta k^{res}\}$ and $\Delta C_i^{com} \in \{\Delta p, \Delta g, \Delta s, \Delta i, \Delta e^{com}, \Delta k^{com}\}$, and satisfies $\Delta C_i^{res}|_{0\to T} \leq 0$ and $\Delta C_i^{com}|_{0\to T} \leq 0$. According to the definition of carbon intensity, the total decarbonization ($DC^{com}$ or $DC^{res}$) can be written as:

$$\begin{cases} DC^{com}|_{0\to T} = DCI^{com}|_{0\to T} \times F|_{0\to T} \\ DC^{res}|_{0\to T} = DCI^{res}|_{0\to T} \times H|_{0\to T} \end{cases} \quad (7)$$

Furthermore, to gain insight into the operational decarbonization potential of residential buildings, the decarbonization efficiency is defined as the ratio of operational carbon emissions to operational decarbonization.

*2.4 Datasets*

The data for the population and economy-related indicators were collected from the World Bank (www.data.worldbank.org). The remaining data related to global building operations were gathered from the Global Building Emissions (GLOBE) Database (http://globe2060.org) [11-15], which was established on the basis of IEA Database.

**3. RESULTS**

*3.1 Global and regional patterns of operational carbon change in residential buildings*

Fig. 1 reveals the global trends and regional patterns of carbon intensity in residential building operations between 2000 and 2020. Overall, the operational carbon emissions from residential buildings of twelve major emitting regions worldwide (all the samples covering 56 countries) increased by 128.8 megatons of carbon dioxide (MtCO$_2$) in the past two decades, but the operational carbon intensity decreased by 764.3 kilograms of carbon dioxide per household (kgCO$_2$/household) with an average decline of 1.2% per yr, and this trend was more pronounced (before 2010: -0.2%, after 2010: -2.2%).

Further details of these regional emission patterns are illustrated in Fig. 1 c. From 2000 to 2020, GDP per capita was the most critical driver increasing the carbon intensity of residential building operations worldwide, which drove the operational carbon intensity of the twelve emitting regions to increase by 56.1% and 26.7% during 2000–2010 and 2010–2020, respectively especially in emerging economies (e.g., China (140.8% and 60.2%) and India (83.6% and 46.9%)). Furthermore, end-use structure was another key to promoting operational carbon intensity. However, energy intensity was the biggest driver in decarbonizing residential building operations, leading to reductions of 20.4%, 23.1%, 19.8%, and 4.3% in carbon intensity during the periods 2000–2005, 2005–2010, 2010–2015, and 2015–2020, respectively, followed by average household size, which contributed to a 3.8%, 2.8%, 2.6%, and 3.2% reduction in carbon intensity during the same periods. Household consumption capacity and emission factors had an unstable effect on regional change in residential carbon intensity.

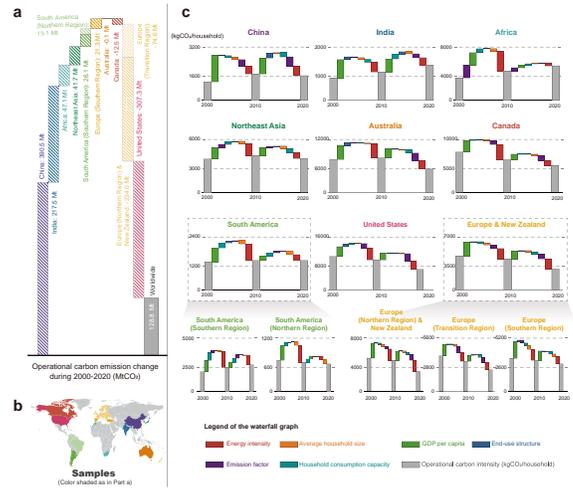

Fig. 1 Global and regional patterns in the carbon emission change of residential building operations.

*3.2 Global and regional patterns of operational carbon change in commercial buildings*

Fig. 2 reveals the change in operational carbon intensity of commercial buildings in the 16 selected countries from 2000 to 2019 based on the assessment of the DSD method. This shows that from 2000 to 2019, the carbon intensity of commercial building operations in all countries except China decreased significantly to varying degrees, with an average decline of -1.94% per yr.

For the drivers of carbon intensity change, during the study period (2000-2019), GDP per capita (global average contribution: 32.22%) was the most significant driver of the increase in the carbon intensity of commercial building operations for all countries, especially in developing economies (e.g., China: 93.51%). Another positive factor is economic efficiency, with an average global contribution of 6.36%. With regard to the drivers that reduce the carbon intensity, emission factors (global average contribution: -75.44%) were the most important driver of carbon intensity reduction in most countries. Meanwhile, besides the United States and Finland, the industrial structure (global average contribution: -20.95%) is another driver to decrease the carbon intensity in most countries. In contrast, the effects of population density and energy intensity on changes in the operational carbon intensity of commercial buildings are not stable.

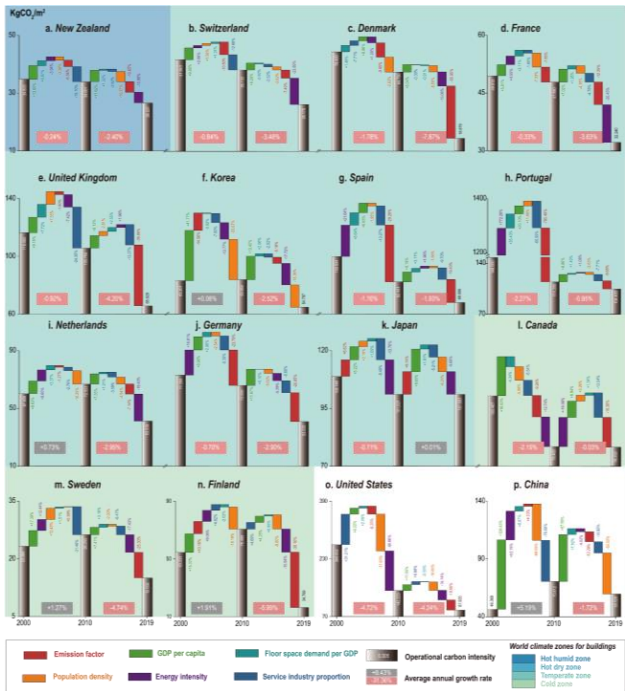

Fig. 2 Changes in the operational carbon intensity of global commercial buildings (2000–2019).

## 4. DISCUSSION

### 4.1 Historical decarbonization of residential building operations

Fig. 3 explores in more detail the global, regional, and national historical processes (i.e., the total cumulative decarbonization) of residential building operations over the past two decades to identify specific regions or countries with higher total decarbonization or decarbonization efficiency that are therefore more likely to be attractive mitigation targets. At the global level (Fig. 3 a-f), the process of decarbonizing residential building operations has been accelerating: by 2020, the cumulative decarbonization in the residential building sector totaled 7.1 GtCO$_2$ worldwide, which is equivalent to nearly nine times the operational carbon emissions of China's residential buildings in 2020. The top three emitters, China, the United States, and India, consistently led the global residential building decarbonization process, contributing, as a whole, more than half of the global decarbonization (54.0%), whereas the remaining 53 countries were distributed between 1.8 MtCO$_2$ (Malta) and 369.4 MtCO$_2$ (Germany). At a regional level (Fig. 3 g-h), from 2000 to 2020, the operational carbon emissions in twelve emitting regions rose rapidly, and the decarbonization efficiency in most regions began to stabilize after 2010. Fig. 3 i analyzes the decarbonization potential of different economies at a national level, and it is worth noting that China and the United States, as the top two emitters, contributed nearly half (48.6%) of the global total decarbonization, whereas the decarbonization efficiency of larger emitters such as India, Japan, Germany, and the United Kingdom were all approximately 6%-9%, which was below the global average level (9.4%). In contrast, some emerging emitters had higher decarbonization potential, for example, Serbia (34.9%), Uruguay (25.3%), Estonia (25.3%), and Ukraine (19.8%), which is expected to become the hot spot in the age of post-COP 27.

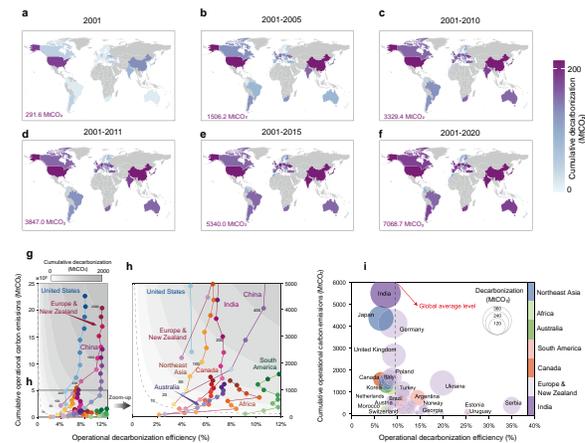

Fig. 3. Global, regional, and national trends in cumulative decarbonization of residential building operations.

Fig. 4 a and b reflect the trend of total decarbonization, the results show that in the past 20 years (2000-2019), the total cumulative decarbonization for 16 countries was 4375.39 MtCO$_2$, which was 10.05% of the total cumulative emissions. In terms of the phased carbon reduction process (as shown in Fig. 4 c), the average share of total cumulative decarbonization over the four periods is 29.08% (2001-2005), 26.71% (2006-2010), 26.81% (2011-2015) and 17.40% (2016-2019), respectively. Of these, the cumulative decarbonization levels of China (1159.90 MtCO$_2$) and the United States (2244.39 MtCO$_2$) were far ahead of other countries in the same period, followed by Germany (236.98 MtCO$_2$), the United Kingdom (161.08 MtCO$_2$) and Korea (159.25 MtCO$_2$), while the cumulative decarbonization of the remaining countries was less than 100 MtCO$_2$ (ranging from 6.91 MtCO$_2$ to 83.84 MtCO$_2$).

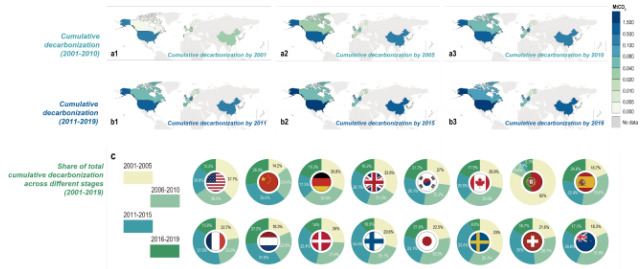

*Fig. 4. Total cumulative decarbonization of global commercial building operations in (a1-a3) 2001-2010 and (b1-b3) 2011-2019; (c) share of the total cumulative decarbonization across different stages.*

*4.2 Historical decarbonization of commercial building operations*

Fig. 5 further analyzes and compares the decarbonization level of residential building operations worldwide at different emission scales (total decarbonization, decarbonization intensity, and decarbonization per capita).

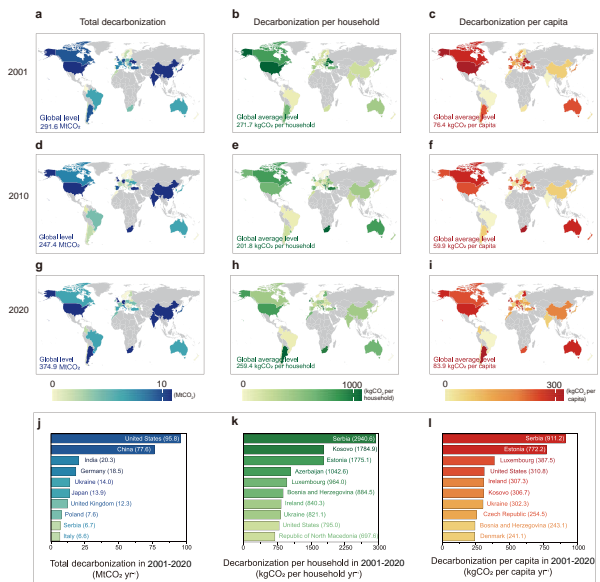

*Fig. 5. Multiscale assessment of the historical decarbonization level of global residential building operations.*

Overall, as shown in Fig. 5 a-i, the assessment results at different emission scales varied significantly and showed remarkable regional heterogeneity. Globally, from 2000 to 2020, although the total decarbonization of residential buildings generally increased, decarbonization per capita and decarbonization per household gradually stabilized or even decreased, and by 2020, the global decarbonization of residential building operations totaled 374.9 MtCO$_2$, and the decarbonization intensity and decarbonization per capita were 259.4 kgCO$_2$/household and 83.9 kgCO$_2$ per capita, respectively. Fig. 5 j-l present the top ten countries with the highest average annual decarbonization levels at three different emission scales and reveal that the countries with higher total decarbonization of residential buildings were not necessarily the ones with higher decarbonization intensity or decarbonization per capita, especially in developing countries. Furthermore, the decarbonization levels of the United States, Ukraine, and Serbia were relatively high at three scales and were all in the top ten, whereas decarbonization levels were poor at all emission scales in less developed economies with high carbon intensity, such as Morocco, Uruguay, and Albania.

Moreover, the change in decarbonization intensity of global commercial operations is shown in Fig. 6 a1-a3. From 2000 to 2019, the gap between the decarbonization intensities of countries is gradually narrowed, with a global decarbonization intensity of 6.86 kgCO$_2$/m$^2$ of commercial building operations in 2019. Fig. 6 b1-b3 shows the dynamics of decarbonization per capita. The per capita decarbonization intensity of global commercial building operations in 2019 was 97.42 kgCO$_2$ per capita, of which the highest was the United States (247.61 kgCO$_2$ per capita), followed by Korea (235.23 kgCO$_2$ per capita), Portugal (208.19 kgCO$_2$ per capita), and Germany (206.14 kgCO$_2$ per capita).

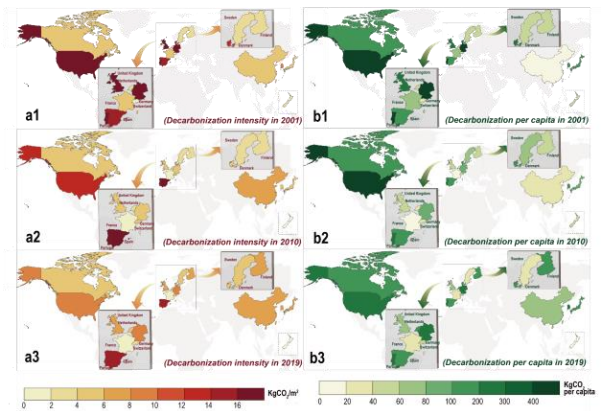

*Fig. 6. Decarbonization intensity and decarbonization per capita for each country in 2001, 2010 and 2019*

## 5. CONCLUSIONS

This study developed a bottom-up modeling framework from the demand side to assess the decarbonization process of global building operations worldwide over the past two decades. First, the DSD approach was integrated to characterize the evolutionary pattern of operational carbon intensity. Then, this work assessed and compared the global, regional, and national historical decarbonization performance and the corresponding potential in global building operations. The key findings are summarized below:

**(1) The operational carbon intensity of global residential buildings continued to decline from 2000 to 2020, dominated by developed regions with higher carbon intensity**. Over the past two decades, the carbon intensity of residential building operations in the twelve emitting regions declined by 764.3 kgCO$_2$/household, with an average annual change of 1.2%. Regarding the drivers behind these regional changes, energy intensity and average household size were the largest contributors decarbonizing global residential building operations, and GDP per capita was the most critical factor driving the increase in carbon emissions in all emitting regions, especially in developing regions.

**(2) The carbon intensity of global commercial building operations continued to decline from 2000 to 2019, and the trend was more significant.** The global carbon intensity of commercial building operations decreased by an average of 1.42% and 2.93% per year in the periods of 2000-2010 and 2010-2019, respectively. GDP per capita as a factor was the most positive factor hindering the decrease in carbon intensity, followed by economic efficiency. On the other hand, the emission factor and industrial structure were key to decarbonizing the commercial building operations.

**(3) The decarbonization process of global residential building operations accelerated, but the opposite is true for global commercial buildings.** For residential buildings, the decarbonization from residential building operations across 56 economies in 2020 was 259.4 kgCO$_2$/household or 83.9 kgCO$_2$ per capita. Globally, residential building operations cumulatively decarbonized 7.1 GtCO$_2$ with a decarbonization efficiency of 9.4%, which is equivalent to approximately nine times the operational carbon of China's residential buildings in 2020, with nearly 80% of this decarbonization coming from the top three emission giants, including United States, China, and Europe & New Zealand. For commercial buildings, from 2000 to 2019, the cumulative decarbonization level of 16 countries was 4375.39 MtCO$_2$, equivalent to 10.05% of the total cumulative emissions, of which 77.81% of decarbonization came from China and the United States.

**ACKNOWLEDGEMENT**

This manuscript has been authored by an author at Lawrence Berkeley National Laboratory under Contract No. DE-AC02-05CH11231 with the U.S. Department of Energy. The U.S. Government retains, and the publisher, by accepting the article for publication, acknowledges, that the U.S. Government retains a non-exclusive, paid-up, irrevocable, world-wide license to publish or reproduce the published form of this manuscript, or allow others to do so, for U.S. Government purposes.